\documentclass[12pt]{article}
\usepackage{graphicx}
\usepackage{amsmath, amssymb}

\newcommand{\half}{{\frac{1}{2}}}
\newcommand{\ket}[1]{\,\left|\,{#1}\right\rangle}
\newcommand{\mbf}[1]{\mathbf{#1}}

\renewcommand{\bar}[1]{\overline{#1}}

\def\Dslash{\raise.15ex\hbox{/}\kern-.7em D}
\def\Pslash{\raise.15ex\hbox{/}\kern-.7em P}

\begin {document}
\begin{flushright}
{\small
SLAC--PUB--13220\\
April 2008}
\end{flushright}

\vspace{30pt}

\centerline{\LARGE \bf Light-Front Holography and}

\vspace{7pt} 

\centerline{\LARGE \bf AdS/QCD Correspondence}

\vspace{20pt}

\centerline{\bf {
Stanley J. Brodsky$^{a}$
 and
Guy F. de T\'eramond$^{b}$}
}
\vspace{15pt}

{\centerline {$^{a}${Stanford Linear Accelerator Center, 
Stanford University, Stanford, CA 94309, USA}}

\vspace{3pt}

{\centerline {$^{b}${Universidad de Costa Rica, San Jos\'e, Costa Rica}}

\vspace{40pt}

\begin{abstract}

Light-Front Holography is a remarkable consequence of the correspondence between string theory in AdS space and conformal field theories in physical-space time. 
It allows string modes $\Phi(z) $ in the AdS fifth dimension to be precisely mapped to the light-front wavefunctions of hadrons in terms of a specific light-front impact variable $\zeta $ which measures the separation of the quark and gluonic constituents within the hadron. This mapping was originally obtained by matching the exact expression for electromagnetic current matrix elements in AdS space with the corresponding exact expression for the current matrix element using light-front theory in physical space-time.  More recently we have shown that one obtains the identical holographic mapping using matrix elements of the energy-momentum tensor, thus providing an important consistency test and verification of holographic mapping from AdS to physical observables defined on the light-front. The resulting light-front Schrodinger equations predicted from AdS/QCD give a good representation of the observed meson and baryon spectra and give excellent phenomenological predictions for amplitudes such as electromagnetic form factors and decay constants.

\end{abstract}

\vspace{25pt}

\begin{center}
{\it Presented at  QCD Down Under II\\
Auckland , New Zealand, January 17--19, 2008
 }
\end{center}

\vfill

\newpage

\section{Introduction}

One of the most challenging problems in strong interaction dynamics is to understand the composition of mesons and
baryons in terms of the fundamental quark and gluon degrees of freedom of the QCD Lagrangian. Because of the strong-coupling of QCD in the infrared domain, it has been difficult to find analytic solutions for the wavefunctions of hadrons or to make precise predictions for hadronic properties outside of the perturbative regime.  Thus an important theoretical goal is to find an initial approximation
to bound-state problems in QCD which is analytically tractable and which can be systematically improved.
Recently 
the AdS/CFT correspondence~\cite{Maldacena:1997re} between string
states in anti--de Sitter (AdS) space and conformal field theories in physical space-time, modified for color confinement,
has led to a semiclassical model for strongly-coupled QCD which provides analytical insights
into its inherently non-perturbative nature, including hadronic spectra, decay constants, and wavefunctions.
As we have recently shown~\cite{Brodsky:2006uqa,Brodsky:2007hb}, there is a remarkable mapping between the AdS description of hadrons and the Hamiltonian formulation of QCD in physical space-time quantized on the light front. 

The natural extension of a wavefunction  for relativistic quantum field theories such as QCD is the light-front wavefunction $\psi_n(x_i, \mbf{k}_{\perp i}, \lambda_i)$ which specifies the  $n$ quark and gluon constituents of a hadron's Fock state as a function of the light-cone fractions $x_i = k^+/P^+ = (k^0+k^z)/(P^0+P^z)$ transverse momenta $\mbf{k}_{\perp i}$ and spin projections $\lambda_i$. The light-front wavefunctions of bound states in QCD are the relativistic
generalizations of the familiar Schr\"odinger wavefunctions of
atomic physics, but they are determined at fixed light-cone time
$\tau  = t +z/c$---the ``front form" advocated by Dirac~\cite{Dirac:1949cp}---rather than
at fixed ordinary time $t$.  
The  light-front wavefunctions  of a
hadron are independent of the momentum of the hadron, and they are
thus boost invariant; Wigner transformations and Melosh rotations
are not required. The light-front formalism for gauge theories in
light-cone gauge is particularly useful in that there are no ghosts,
and one has a direct physical interpretation of  orbital angular
momentum.

When a flash from a camera illuminates a scene, each object is illuminated along the light-front of the flash; i.e., at a given $\tau$.  Similarly, when a sample is illuminated by an x-ray source such as the Linac Coherent Light  Source, each element of the target is struck at a given $\tau.$  In contrast, setting the initial condition using conventional instant time $t$ requires simultaneous scattering of photons on each constituent. 
Thus it is natural to set boundary conditions at fixed $\tau$ and then evolve using the light-front Hamiltonian $P^- = P^0-P^z = i {d/d \tau}.$  The invariant Hamiltonian $H_{LF} = P^+ P^- - P^2_\perp$ then has eigenvalues $\mathcal{M}^2$ where $\mathcal{M}$ is the physical mass.   Its eigenfunctions are the light-front eigenstates whose Fock state projections define the light-front wavefunctions.

{\it Light-Front Holography} is an important feature of AdS/CFT; it allows string modes $\Phi(z)$ in the AdS fifth dimension
to be precisely mapped to the light-front wavefunctions of hadrons in physical space-time in
terms of a specific light-front impact variable $\zeta$ which measures the separation of the
quark and gluonic  constituents  within the hadron.
The AdS/CFT correspondence implies
that a strongly coupled gauge theory is equivalent to the propagation
of weakly coupled strings in a higher dimensional space,
where physical quantities are computed in terms of an effective
gravitational theory. Thus, the AdS/CFT duality provides a gravity
description in a ($d+1$)-dimensional AdS
space-time in terms of a
$d$-dimensional conformally-invariant quantum field theory at the AdS 
asymptotic boundary~\cite{Gubser:1998bc,Witten:1998qj}.

QCD is a confining theory in the infrared
with a mass gap $\Lambda_{\rm QCD}$ and a well-defined spectrum of color singlet states.
Conformal symmetry is broken in physical QCD by quantum effects and quark masses.
There are indications however,  both from
theory and phenomenology, that the QCD coupling is slowly varying at small momentum 
transfer~\cite{Brodsky:2008pg}. In particular, a new extraction of the effective strong coupling constant
$\alpha_{s,g_1}(Q^2)$
from CLAS spin structure function data in an extended $Q^2$ region
using the Bjorken sum $\Gamma_1^{p-n}(Q^2)$~\cite{Deur:2008rf},
indicates the lack of $Q^2$ dependence of $\alpha_s$ in the low $Q^2$ limit. One can understand this
physically~\cite{Brodsky:2008pg}: in a confining theory where gluons have an effective 
mass or maximal wavelength, all vacuum polarization
corrections to the gluon self-energy decouple at long wavelength; thus an infrared fixed point appears to be a natural consequence of confinement~\cite{Cornwall:1981zr}.  Furthermore, if one considers a
semi-classical approximation to QCD with massless quarks and without
particle creation or absorption, then the resulting $\beta$ function
is zero, the coupling is constant, and the approximate theory is
scale and conformal invariant~\cite{Parisi:1972zy}. One can thus use conformal symmetry as 
a {\it template}, systematically correcting for its nonzero $\beta$ function as well
as higher-twist effects~\cite{Brodsky:1985ve}.

Different values of the holographic variable $z$ determine the scale of the invariant
separation between the partonic constituents. 
Hard scattering processes occur in the small-$z$ ultraviolet (UV)
region of AdS space. In particular,
the $Q \to \infty$ zero separation limit corresponds to the $z \to 0$ asymptotic boundary, where the QCD
Lagrangian is defined. 
In the large-$z$ infrared (IR) region  a cut-off is introduced to truncate the regime where the AdS modes can propagate. The infrared cut-off breaks conformal invariance, allows the introduction of
a scale and a spectrum of particle states. In the hard wall model~\cite{Polchinski:2001tt}
 a cut-off is placed at a
finite value $z_0 = 1/\Lambda_{\rm QCD}$ and the spectrum of states is linear in the radial and angular momentum quantum numbers:
$\mathcal{M} \sim 2 n \! + \! L$. In the soft wall model a smooth infrared cutoff is chosen to model
confinement and reproduce the usual Regge behavior 
$\mathcal{M}^2 \sim n \! + \!  L$~\cite{Karch:2006pv}.
The resulting models, although {\it ad hoc}, provide a
simple semi-classical approximation to QCD which has both constituent counting
rule behavior at short distances and confinement at large distances~\cite{Brodsky:2008pg}.

It is thus natural, as a useful first approximation, to use the isometries of AdS to map the local interpolating operators at the UV boundary of AdS space to the modes propagating inside AdS.
The short-distance behavior of a hadronic state is
characterized by its twist  (dimension minus spin) 
$\tau = \Delta - \sigma$, where $\sigma$ is the sum over the constituent's spin
$\sigma = \sum_{i = 1}^n \sigma_i$. Twist is also equal to the number of partons $\tau = n$.
Under conformal transformations the interpolating operators transform according to their twist, and consequently the AdS isometries map the twist scaling dimensions into the AdS 
modes~\cite{Brodsky:2003px}.

The eigenvalues of normalizable modes
in AdS give the hadronic spectrum. AdS modes represent also the probability
amplitude for the distribution of quarks and gluons at a given scale.
There are also non-normalizable modes which are related to 
external currents: they propagate into the AdS  interior  and couple to
boundary QCD interpolating operators~\cite{Gubser:1998bc,Witten:1998qj}.
Following this simplified ``bottom up" approach,  a limited set of operators is  introduced to construct 
phenomenological viable five-dimensional dual holographic 
models~\cite{Boschi-Filho:2002vd,deTeramond:2005su,Erlich:2005qh,DaRold:2005zs}.

An important feature of light-front
quantization is the fact that it provides exact formulas for
current matrix elements as a sum of bilinear forms which can be mapped
into their AdS/CFT counterparts in the semi-classical approximation.
The AdS metric written in terms of light front coordinates $x^\pm =
x^0 \pm x^3$ is
\begin{equation} \label{eq:AdSzLF}
ds^2 = \frac{R^2}{z^2} \left( dx^+ dx^- - d \mbf{x}_\perp^2 - dz^2
\right).
\end{equation}
At fixed light-front time $x^+=0$, the metric depends only on the transverse
$ \mbf{x}_\perp$ and the holographic variable $z$.
Thus we can find an exact correspondence between the
fifth-dimensional coordinate of anti-de Sitter space $z$ and a
specific impact variable $\zeta$ in the light-front formalism.  The
new variable $\zeta$
measures the separation of the constituents within the hadron in
ordinary space-time.  The amplitude $\Phi(z)$ describing  the
hadronic state in $\rm{AdS}_5$ can then be precisely mapped to the valence
light-front wavefunctions $\psi_{n/H}$ of hadrons in physical
space-time~\cite{Brodsky:2006uqa,Brodsky:2007hb}, thus providing a relativistic
description of hadrons in QCD at the amplitude level.

The correspondence of AdS amplitudes to the QCD wavefunctions in light-front coordinates
was carried out in~\cite{Brodsky:2006uqa,Brodsky:2007hb}
by comparing the expressions for the electromagnetic matrix elements in QCD and
AdS for any value of the momentum transfer $q^2$.   More recently we have shown that one obtains the identical holographic mapping using the matrix elements of the energy-momentum 
tensor~\cite{Brodsky:2008pf}. To prove this, we show that there exists a correspondence between the matrix elements of the energy-momentum tensor  of the 
fundamental hadronic constituents in QCD with the transition amplitudes
describing the interaction of string modes in
anti-de Sitter space with an external graviton field which propagates in the AdS interior.  The proof is outlined below.  

The matrix elements of the energy-momentum tensor $\Theta^{\mu \nu}$ of each constituent define the gravitational form factor of a composite hadron.
One can also use gravitational matrix elements to obtain the holographic mapping of the AdS mode  wavefunctions $\Phi(z)$ in  AdS space to the light-front wavefunctions 
$\psi_{n/H}$ in physical $3+1$ space-time  defined at fixed light-cone time $\tau= t + z/c$.   We find the identical holographic mapping from $z \to \zeta$ as in the electromagnetic case. The agreement of the results for electromagnetic and gravitational hadronic transition amplitudes provides an important consistency test and verification of holographic mapping from AdS to physical observables defined on the light-front. 
It is indeed remarkable that 
such a correspondence exists, since strings describe extended objects
coupled to an electromagnetic field distributed in the AdS interior, whereas QCD
degrees of freedom are pointlike particles with individual local couplings to the electromagnetic
current. However, as we have recently shown~\cite{Brodsky:2006uqa,Brodsky:2007hb,Brodsky:2008pf}, a precise mapping of AdS modes to hadronic light-front wavefunctions
can be found in the strongly coupled semiclassical approximation to QCD.

\section{The Light-Front Fock Representation}

The light-front expansion of any hadronic system
is constructed by quantizing QCD
at fixed light-cone time~\cite{Dirac:1949cp} $\tau = t + z/c$.
In terms of the hadron  four-momentum $P = (P^+, P^-, \mbf{P}_{\!\perp})$,
$P^\pm = P^0 \pm P^3$,
the light-cone Lorentz invariant Hamiltonian for the composite system, 
$H_{LF}^{QCD} = P^-P^+ - \mbf{P}^2_\perp$,  has
eigenvalues given in terms of the eigenmass ${\cal M}$ squared  corresponding 
to the mass spectrum of the color-singlet states in QCD~\cite{Brodsky:1997de}
$H_{LF} \vert{\psi_H}\rangle = {\cal M}^2_H \vert{\psi_H}\rangle$.
Each hadronic eigenstate $\vert \psi_H \rangle$  can be expanded in
a Fock-state complete basis of non-interacting $n$-particle states
$\vert n \rangle$ with an infinite number of components
\begin{multline}
\left\vert \psi_H(P^+,\mbf{P}_{\! \perp}, S_z) \right\rangle = 
\sum_{n,\lambda_i} 
\prod_{i=1}^n \int \! \frac{dx_i}{\sqrt{x_i}}
\frac{d^2 \mbf{k}_{\perp i}}{2 (2\pi)^3} \, 16 \pi^3 \,
\delta \Bigl(1 - \sum_{j=1}^n x_j\Bigr) 
\delta^{(2)} \negthinspace\Bigl(\sum_{j=1}^n\mbf{k}_{\perp j}\Bigr) \\
\times \psi_{n/H}(x_i,\mbf{k}_{\perp i},\lambda_i) 
\bigl\vert n: x_i P^+\!, x_i \mbf{P}_{\! \perp} \! + \! \mbf{k}_{\perp i},\lambda_i \bigr\rangle,
\label{eq:LFWFexp}
\end{multline}
where the sum begins with the valence state; e.g., $n \ge 3$ for baryons. The
coefficients of the  Fock expansion
$\psi_{n/H}(x_i, \mbf{k}_{\perp i},\lambda_i) 
= \bigl\langle n:x_i,\mbf{k}_{\perp i},\lambda_i \big\vert \psi_H\bigr\rangle$ ,
are independent of the total momentum $P^+$ and $\mbf{P}_{\! \perp}$ of
the hadron and depend only on the relative partonic coordinates,
the longitudinal momentum fraction $x_i = k_i^+/P^+$,
the relative transverse momentum $\mbf{k}_{\perp i}$
and $\lambda_i$, the
projection of the constituent's spin along the $z$ direction. 
Momentum conservation requires
$\sum_{i=1}^n x_i =1$ and $\sum_{i=1}^n \mbf{k}_{\perp i}=0$.
In addition, each light front wavefunction
$\psi_{n/H}(x_i,\mbf{k}_{\perp i},\lambda_i)$ obeys the angular momentum sum 
rule~\cite{Brodsky:2000ii}
$J^z  = \sum_{i=1}^n  S^z_i + \sum_{i=1}^{n-1} L^z_i $,
where $S^z_i = \lambda_i $ and the $n-1$ orbital angular momenta
have the operator form 
$L^z_i =-i \left(\frac{\partial}{\partial k^x_i}k^y_i -
\frac{\partial}{\partial k^y_i}k^x_i \right)$.
It should be emphasized that the assignment of quark and gluon spin and orbital angular momentum of a hadron is  a gauge-dependent concept. The LF framework in light-cone gauge $A^+=0$ provides a 
physical definition since there are no gauge field ghosts and the gluon 
has spin-projection $J^z= \pm 1$; moreover, it is frame-independent.

\section{Light-Front Holography}

Light-Front Holography can be derived by observing the correspondence between matrix elements obtained in AdS/CFT with the corresponding formula using the LF representation.  We will outline the basic elements of the derivation in this section.
For simplicity  we discuss the specific mapping for two-parton hadronic state. The $n$-parton
case is derived~\cite{Brodsky:2008pf} using an effective single particle 
density~\cite{Soper:1976jc}.

We first consider the light-front electromagnetic form factor in impact 
space~\cite{Brodsky:2006uqa,Brodsky:2007hb}
\begin{equation} \label{eq:FFb} 
F(q^2) =  \sum_n  \prod_{j=1}^{n-1}\int d x_j d^2 \mbf{b}_{\perp j}  \sum_q e_q
\exp \! {\Bigl(i \mbf{q}_\perp \! \cdot \sum_{j=1}^{n-1} x_j \mbf{b}_{\perp j}\Bigr)} 
\left\vert \tilde \psi_n(x_j, \mbf{b}_{\perp j})\right\vert^2,
\end{equation}
written as a sum of overlap of light-front wave functions of the $j = 1,2, \cdots, n-1$ spectator
constituents.  We have included
explicitly in (\ref{eq:FFb})
the contribution from each active constituent $q$ with charge $e_q$.
The formula is exact if the sum is over all Fock states $n$.

For definiteness we shall consider a two-quark $\pi^+$  valence Fock state 
$\vert u \bar d\rangle$ with charges $e_u = \frac{2}{3}$ and $e_{\bar d} = \frac{1}{3}$.
For $n=2$, there are two terms which contribute to the $q$-sum in (\ref{eq:FFb}). 
Exchanging $x \leftrightarrow 1-x$ in the second integral  we find ($e_u + e_{\bar d}$ = 1)
\begin{eqnarray} \nonumber
 F_{\pi^+}(q^2)  &=& \!  \int_0^1 \! d x \int \! d^2 \mbf{b}_{\perp}  
 e^{i \mbf{q}_\perp \cdot  \mbf{b}_{\perp} (1-x)} 
\left\vert \tilde \psi_{u \bar d/ \pi}\! \left(x,  \mbf{b}_{\perp }\right)\right\vert^2 \\
\label{eq:PiFFb}
&=&  2 \pi \int_0^1 \! \frac{dx}{x(1-x)}  \int \zeta d \zeta\, 
J_0 \! \left(\! \zeta q \sqrt{\frac{1-x}{x}}\right) 
\left\vert\tilde\psi_{u \bar d/ \pi}\!(x,\zeta)\right\vert^2,
\end{eqnarray}
where $\zeta^2 =  x(1-x) \mathbf{b}_\perp^2$ and $F_\pi^+(q\!=\!0)=1$. 
Notice that by performing an identical calculation for the
$\pi^0$ meson the result is $F_{\pi^0}(q^2) = 0$ for any $q$, as expected
from $C$-charge conjugation invariance.

We now compare this result with the electromagnetic form-factor in AdS space:
\begin{equation} 
F(Q^2) = R^3 \int \frac{dz}{z^3} \, J(Q^2, z) \vert \Phi(z) \vert^2,
\label{eq:FFAdS}
\end{equation}
where $J(Q^2, z) = z Q K_1(z Q)$.
Using the integral representation of $J(Q^2,z)$
\begin{equation} \label{eq:intJ}
J(Q^2, z) = \int_0^1 \! dx \, J_0\negthinspace \left(\negthinspace\zeta Q
\sqrt{\frac{1-x}{x}}\right) ,
\end{equation} we can write the AdS electromagnetic form-factor as
\begin{equation} 
F(Q^2)  =    R^3 \! \int_0^1 \! dx  \! \int \frac{dz}{z^3} \, 
J_0\!\left(\!z Q\sqrt{\frac{1-x}{x}}\right) \left \vert\Phi(z) \right\vert^2 .
\label{eq:AdSFx}
\end{equation}
Comparing with the light-front QCD  form factor (\ref{eq:PiFFb}) for arbitrary  values of $Q$
\begin{equation} \label{eq:Phipsi} 
\vert \tilde\psi(x,\zeta)\vert^2 = 
\frac{R^3}{2 \pi} \, x(1-x)
\frac{\vert \Phi(\zeta)\vert^2}{\zeta^4}, 
\end{equation}
where we identify the transverse light-front variable $\zeta$, $0 \leq \zeta \leq \Lambda_{\rm QCD}$,
with the holographic variable $z$.

Matrix elements of the energy-momentum tensor $\Theta^{\mu \nu} $ which define the gravitational form factors play an important role in hadron physics.  Since one can define $\Theta^{\mu \nu}$ for each parton, one can identify the momentum fraction and  contribution to the orbital angular momentum of each quark flavor and gluon of a hadron. For example, the spin-flip form factor $B(q^2)$ which is the analog of the Pauli form factor $F_2(Q^2)$ of a nucleon provides a  measure of the orbital angular momentum carried by each quark and gluon constituent of a hadron at $q^2=0.$   Similarly,  the spin-conserving form factor $A(q^2)$, the analog of the Dirac form factor $F_1(q^2)$, allows one to measure the momentum  fractions carried by each constituent.
This is the underlying physics of Ji's sum rule~\cite{Ji:1996ek}:
$\langle J^z\rangle = \half [ A(0) + B(0)] $,  which has prompted much of the current interest in 
the generalized parton distributions (GPDs)  measured in deeply
virtual Compton scattering. Measurements of the GDP's are of particular relevance
for determining the distribution of partons in the transverse
impact plane, and thus could be confronted with AdS/QCD predictions which follow
from the mapping of AdS modes to the transverse impact representation~\cite{Brodsky:2006uqa}.

An important constraint is $B(0) = \sum_i B_i(0) = 0$;  i.e.  the anomalous gravitomagnetic moment of a hadron vanishes when summed over all the constituents $i$. This was originally derived from the equivalence principle of gravity~\cite{Teryaev:1999su}.  The explicit verification of these relations, Fock state by Fock state, can be obtained in the light-front quantization of QCD in  light-cone 
gauge~\cite{Brodsky:2000ii}.  Physically $B(0) =0$ corresponds to the fact that the sum of the $n$ orbital angular momenta $L$ in an $n$-parton Fock state must vanish since there are only $n-1$ independent orbital angular momenta.

Gravitational form factors can also be computed in AdS/QCD from the overlap integral of
hadronic string modes propagating in AdS space with a graviton field $h_{\mu \nu}$ 
which acts as a source and probes the AdS interior. This has  been done very recently for the gravitational form factors of mesons  by Abidin and Carlson~\cite{Abidin:2008ku}, thus providing restrictions on the GPDs.

The light-front expression for the helicity-conserving gravitational form factor in impact space
is~\cite{Brodsky:2008pf}
\begin{equation} \label{eq:Ab}
A(q^2) =  \sum_n  \prod_{j=1}^{n-1}\int d x_j d^2 \mbf{b}_{\perp j}  \sum_f x_f
\exp \! {\Bigl(i \mbf{q}_\perp \! \cdot \sum_{j=1}^{n-1} x_j \mbf{b}_{\perp j}\Bigr)} 
\left\vert \tilde \psi_n(x_j, \mbf{b}_{\perp j})\right\vert^2,
\end{equation}
which includes the contribution of each struck parton with longitudinal momentum $x_f$
and corresponds to a change of transverse momentum $x_j \mbf{q}$ for
each of the $j = 1, 2, \cdots, n-1$ spectators. 
For $n=2$, there are two terms which contribute to the $f$-sum in  (\ref{eq:Ab}). 
Exchanging $x \leftrightarrow 1-x$ in the second integral we find 
\begin{eqnarray} \label{eq:PiGFFb} \nonumber
A_{\pi}(q^2) &\! = \!&  2 \! \int_0^1 \! x \, d x \int \! d^2 \mbf{b}_{\perp}  
 e^{i \mbf{q}_\perp \cdot  \mbf{b}_{\perp} (1-x)} 
\left\vert \tilde \psi \left(x, \mbf{b}_{\perp }\right)\right\vert^2 \\
&\! = \!& 4 \pi \int_0^1 \frac{dx}{(1-x)}  \int \zeta d \zeta\,
J_0 \! \left(\! \zeta q \sqrt{\frac{1-x}{x}}\right)  \vert\tilde\psi(x,\zeta)\vert^2,
\end{eqnarray}
where $\zeta^2 =  x(1-x) \mathbf{b}_\perp^2$.  It is simple to prove that if 
$\psi$ is a symmetric function of $x$ and  $(1-x)$ then
\begin{equation}
\int_0^1 \! x \, dx  \int \! d^2 \mbf{b}_{\perp}  \vert \tilde \psi(x, \mbf{b}_\perp) \vert^2 = \frac{1}{2},
\end{equation}
and thus $A_\pi(q^2)$ satisfy the sum rule $A_\pi(q\!=\!0) = 1$.

We now consider the expression for the hadronic gravitational form factor in AdS space
\begin{equation} 
A_\pi(Q^2)  =  R^3 \! \! \int \frac{dz}{z^3} \, H(Q^2, z) \left\vert\Phi_\pi(z) \right\vert^2,
\end{equation}
where $H(Q^2, z) = \half  Q^2 z^2  K_2(z Q)$.
The hadronic form factor is normalized to one at $Q=0$, $A(0) = 1$.
Using the integral representation of $H(Q^2,z)$
\begin{equation} \label{eq:intHz}
H(Q^2, z) =  2  \int_0^1\!  x \, dx \, J_0\!\left(\!z Q\sqrt{\frac{1-x}{x}}\right) ,
\end{equation}
we can write the AdS gravitational form factor 
\begin{equation} 
A(Q^2)  =  2  R^3 \! \int_0^1 \! x \, dx  \! \int \frac{dz}{z^3} \, 
J_0\!\left(\!z Q\sqrt{\frac{1-x}{x}}\right) \left \vert\Phi(z) \right\vert^2 .
\label{eq:AdSAx}
\end{equation}
Comparing with the QCD  gravitational form factor (\ref{eq:PiGFFb}) we find an  identical  relation between the light-front wave function $\tilde\psi(x,\zeta)$ and the AdS wavefunction $\Phi(z)$
in Eq. (\ref{eq:Phipsi}) obtained from the mapping of the pion electromagnetic transition amplitude.

\section{Holographic Light-Front Hamiltonian and Schr\"odinger Equation}

The above analysis provides an exact correspondence between the holographic variable $z$ and an
impact variable $\zeta$ which measures the transverse separation of the constituents within
a hadron, we can identify $\zeta = z$. The mapping of $z$ from AdS space to $\zeta$ in light-front frame  allows the equations of motion in AdS space to be recast in the form of  a
light-front Hamiltonian equation~\cite{Brodsky:1997de}
\begin{equation}
H_{LF} \ket{\phi} = \mathcal{M}^2 \ket{\phi}, \label{eq:HLC}
\end{equation}
a remarkable result which allows the discussion of the AdS/CFT solutions in terms of light-front equations in physical 3+1 space time.
By substituting $\phi(\zeta) =
\zeta^{-3/2} \Phi(\zeta)$ in the AdS scalar wave
equation 
we find an effective Schr\"odinger equation as a function of the
weighted impact variable $\zeta$~\cite{Brodsky:2006uqa,Brodsky:2007hb}
\begin{equation} \label{eq:Scheq}
\left[-\frac{d^2}{d \zeta^2} + V(\zeta) \right] \phi(\zeta) =
\mathcal{M}^2 \phi(\zeta),
\end{equation}
with the conformal potential $V(\zeta) \to - (1-4 L^2)/4\zeta^2$,
an effective two-particle light-front radial equation for mesons.
Its eigenmodes determine the hadronic mass spectrum. 
We have written above $(\mu R)^2 = - 4 + L^2$. 
The holographic hadronic light-front
wave functions $\phi(\zeta) = \langle \zeta \vert \phi \rangle$ are
normalized according to
\begin{equation}
\langle \phi \vert \phi \rangle = \int d\zeta \, \vert \langle
\zeta \vert \phi \rangle \vert^2 = 1,
\end{equation}
and represent the probability amplitude to find $n$-partons at
transverse impact separation $\zeta = z$.   Its
eigenvalues are determined by the boundary conditions at 
$\phi(z =1/\Lambda_{\rm QCD}) = 0$ and are given in terms of the roots of
Bessel functions: $\mathcal{M}_{L,k} = \beta_{L,k} \Lambda_{\rm
QCD}$. The normalizable modes are
\begin{equation}
\phi_{L,k}( \zeta) =   \frac{ \sqrt{2} \Lambda_{\rm QCD}}{J_{1+L}(\beta_{L,k})}
 \sqrt{\zeta} J_L \! \left(\zeta \beta_{L,k} \Lambda_{\rm QCD}\right)
 \theta\big(\zeta \le
\Lambda^{-1}_{\rm QCD}\big).
\end{equation}

The lowest stable state $L = 0$ is determined by the
Breitenlohner-Freedman bound~\cite{Breitenlohner:1982jf}.
Higher excitations are matched to the small $z$ asymptotic behavior of each string mode
to the corresponding
conformal dimension of the boundary operators
of each hadronic state. The effective wave equation
(\ref{eq:Scheq}) is a relativistic light-front equation defined at
$x^+ = 0$. The AdS metric $ds^2$ (\ref{eq:AdSzLF})  is invariant if
$\mbf{x}_\perp^2 \to \lambda^2 \mbf{x}_\perp^2$ and $z \to \lambda
z$ at equal light-front time  $x^+ = 0$. The Casimir operator for the rotation
group $SO(2)$ in the transverse light-front plane is $L^2$. This
shows the natural holographic connection to the light front.

For higher spin-$S$ bosonic modes~\cite{l'Yi:1998eu}, we can also recast the AdS wave 
equation describing a $p$-form field $\Phi_P$ ($P = S$) 
into its light-front form (\ref{eq:HLC}). Using the substitution  
$\phi_S(\zeta) = \zeta^{-3/2+S} \Phi_S(\zeta)$, $\zeta = z$, we find a light-front Schr\"odinger equation
identical to (\ref{eq:Scheq}) with $\phi  \to \phi_S$,
provided that $(\mu R)^2 = - (2-S)^2 + L^2$. 
Stable solutions satisfy a generalized Breitenlohner-Freedman bound
$(\mu R)^2 \ge -  (d - 2 S)^2/4$,
and thus the lowest stable state has scaling dimensions $\Delta = 2$,
 independent of $S$.
The fundamental light-front equation of AdS/CFT has the appearance of a
Schr\"odinger  equation, but it is relativistic, covariant, and analytically tractable.

The pseudoscalar meson interpolating operator
$\mathcal{O}_{2+L}= \bar q \gamma_5 D_{\{\ell_1} \cdots D_{\ell_m\}} q$, 
written in terms of the symmetrized product of covariant
derivatives $D$ with total internal space-time orbital
momentum $L = \sum_{i=1}^m \ell_i$, is a twist-two, dimension $3 + L$ operator
with scaling behavior determined by its twist-dimension $ 2 + L$. Likewise
the vector-meson operator
$\mathcal{O}_{2+L}^\mu = \bar q \gamma^\mu D_{\{\ell_1} \cdots D_{\ell_m\}} q$
has scaling dimension $2 + L$.  The scaling behavior of the scalar and vector AdS modes is precisely the scaling required to match the scaling dimension of the local pseudoscalar and vector-meson interpolating operators.    The spectral predictions for the hard wall model for both light meson and baryon states is compared with experimental data in~\cite{Brodsky:2008pg}.

A closed form of the light-front wavefunctions $\tilde\psi(x, \mbf{b}_\perp)$ follows from
(\ref{eq:Phipsi}) 
\begin{multline} 
\tilde \psi_{L,k}(x, \mbf{b}_\perp) 
=  \frac {\Lambda_{\rm QCD}}{\sqrt{ \pi} J_{1+L}(\beta_{L,k})} \sqrt{x(1-x)} \\ \times
J_L \! \left(\sqrt{x(1-x)} \, \vert\mbf{b}_\perp\vert \beta_{L,k} \Lambda_{\rm QCD}\right) 
\theta \! \left(\mbf{b}_\perp^2 \le \frac{\Lambda^{-2}_{\rm QCD}}{x(1-x)}\right).
\end{multline}
The resulting wavefunction 
displays confinement at large interquark
separation and conformal symmetry at short distances, reproducing dimensional counting rules for hard exclusive amplitudes.

\section{Conclusions}

Light-Front Holography is one of the most remarkable features of AdS/CFT.  It  allows one to project the functional dependence of the wavefunction $\Phi(z)$ computed  in the  AdS fifth dimension to the  hadronic frame-independent light-front wavefunction $\psi(x_i, \mbf{b}_{\perp i})$ in $3+1$ physical space-time. The 
variable $z $ maps  to $ \zeta(x_i, \mbf{b}_{\perp i})$.   As we have discussed, this correspondence is a consequence of the fact that the metric $ds^2$ for AdS$_5$ at fixed light-front time $\tau$ is invariant under the simultaneous scale change  $\mbf{x}^2_\perp \to \lambda^2 \mbf{x}^2_\perp $ in transverse space and $z^2 \to \lambda^2 z^2$.  The transverse coordinate $\zeta$ is closely related to the invariant mass squared  of the constituents in the LFWF  and its off-shellness  in  the light-front kinetic energy,  and it is thus the natural variable to characterize the hadronic wavefunction.  In fact $\zeta$ is the only variable to appear in the light-front
Schr\"odinger equations predicted from AdS/QCD.  These equations for both meson and baryons give a good representation of the observed hadronic spectrum, especially in the case of the soft wall model. The resulting LFWFs also have excellent phenomenological features, including predictions for the  electromagnetic form factors and decay constants.  

It is interesting to note that the form of the nonperturbative pion distribution amplitude $ \phi_\pi(x)$ obtained from integrating the $ q \bar q$ valence LFWF $\psi(x, \mbf{k}_\perp)$  over $\mbf{k}_\perp$,
has a quite different $x$-behavior than the
asymptotic distribution amplitude predicted from the PQCD
evolution~\cite{Lepage:1979zb} of the pion distribution amplitude.
The AdS prediction
$ \phi_\pi(x)  = \sqrt{3}  f_\pi \sqrt{x(1-x)}$ has a broader distribution than expected from solving the Efremov-Radyushkin-Brodsky-Lepage (ERBL) evolution equation in perturbative QCD.
This observation appears to be consistent with the results of the Fermilab diffractive dijet 
experiment~\cite{Aitala:2000hb}, the moments obtained from lattice QCD~\cite{Brodsky:2008pg} and pion form factor data~\cite{Choi:2006ha}.

Nonzero quark masses are naturally incorporated into the AdS predictions~\cite{Brodsky:2008pg} by including them explicitly in the LF kinetic energy  $\sum_i  \frac{\mbf{k}^2_{\perp i} + m_i^2}{x_i}$. Given the nonpertubative LFWFs one can predict many interesting phenomenological quantities such as heavy quark decays, generalized parton distributions and parton structure functions.  
The AdS/QCD model is semi-classical and thus only predicts the lowest valence Fock state structure of the hadron LFWF.  In principle, the model can be systematically improved by diagonalizing the full QCD light-front Hamiltonian on the AdS/QCD basis.

Another interesting application is hadronization at the amplitude level.  In this case one uses light-front time-ordered perturbation theory for the QCD light-front Hamiltonian to generate the off-shell  quark and gluon T-matrix helicity amplitudes such as $e^+ e^- \to \gamma^* \to X$. The amplitude can be renormalized using the ``alternate denominator" method~\cite{Brodsky:1973kb}.
If at any stage a set of  color-singlet partons has  light-front kinetic energy 
$\sum_i {\mbf{k}^2_{\perp i}/ x_i} < \Lambda^2_{QCD}$, then one coalesces the virtual partons into a hadron state using the AdS/QCD LFWFs. A similar approach was used to predict antihydrogen formation from virtual positron--antiproton states produced in $\bar p A$ 
collisions~\cite{Munger:1993kq}.

The hard wall AdS/QCD model resembles  bag models where a boundary condition is introduced to implement confinement.  However, unlike traditional bag models, the AdS/QCD is frame-independent.  An important property of bag models is the  dominance of quark interchange as the underlying dynamics of large-angle elastic scattering,  This agrees with the survey of 
two-hadron exclusive reactions~\cite{White:1994tj}.  In addition the AdS/QCD model implies a maximal wavelength for confined quarks and gluons and thus a finite IR fixed point for the QCD coupling. It is also consistent with the picture that QCD condensates only appear within the 
hadronic boundary~\cite{Brodsky:2008xm}.

Our analysis also allows one to predict the individual quark and gluon contributions to the gravitational form factors $A(q^2)$ and $B(q^2).$  Thus we can immediately predict the momentum fractions 
for quarks $q$ and gluons $g$, $A_{q,g}(0) =  \langle x_{q,g}\rangle$, and orbital angular momenta $B_{q,g}(0) = \langle L_{q,g} \rangle$ carried by each quark flavor and gluon in the hadron with sum rules $\sum_{q,g}A_{q,g}(0) = A(0)= 1$ and $\sum_{q,g} B_{q,g}(0) =B(0) = 0$.  The last sum rule corresponds to the vanishing of the  anomalous gravitational moment which is true Fock state by 
Fock state~\cite{Brodsky:2000ii} in light-front theory. 

The holographic mapping from $\Phi(z)$ to the light-front wave functions of relativistic composite systems provides a new tool for extending the AdS/CFT correspondence to theories such as QCD which are not conformally invariant.
The mathematical consistency of light-front holography for both the electromagnetic and gravitational hadronic transition matrix elements demonstrates that the mapping between the AdS holographic  variable $z$ and the transverse light-front variable $\zeta,$ which is a function of the multi-dimensional coordinates of the partons in a given light-front Fock state $x_i, \mbf{b}_{\perp i}$ at fixed light-front time $\tau,$ is a general principle.

\section*{Acknowledgments}
Presented by SJB at the workshop, QCD Down Under II, 
at Massey University in Auckland, New Zealand,  January 17-19, 2008.  He thanks Tony Signal and Patrick Bowman for organizing this meeting and for their outstanding hospitality.
This research was supported by the Department
of Energy contract DE--AC02--76SF00515. SLAC-PUB-13220.


\begin{thebibliography}{0}

\bibitem{Maldacena:1997re}
  J.~M.~Maldacena,
  ``The large N limit of superconformal field theories and supergravity,''
  Adv.\ Theor.\ Math.\ Phys.\  {\bf 2}, 231 (1998)
  [Int.\ J.\ Theor.\ Phys.\  {\bf 38}, 1113 (1999)]
  [arXiv:hep-th/9711200].
  
      \bibitem{Brodsky:2006uqa}
  S.~J.~Brodsky and G.~F.~de Teramond,
  ``Hadronic spectra and light-front wavefunctions in holographic QCD,''
  Phys.\ Rev.\ Lett.\  {\bf 96}, 201601 (2006)
  [arXiv:hep-ph/0602252].
  
    \bibitem{Brodsky:2007hb}
  S.~J.~Brodsky and G.~F.~de Teramond,
  ``Light-Front Dynamics and AdS/QCD Correspondence: The Pion Form Factor in the Space- and
  Time-Like Regions,''
  Phys.\ Rev.\  D {\bf 77}, 056007 (2008),
  arXiv:0707.3859 [hep-ph].
  
  \bibitem{Dirac:1949cp}
  P.~A.~M.~Dirac,
  ``Forms of Relativistic Dynamics,''
  Rev.\ Mod.\ Phys.\  {\bf 21}, 392 (1949).
  
    
   \bibitem{Gubser:1998bc}
  S.~S.~Gubser, I.~R.~Klebanov and A.~M.~Polyakov,
  ``Gauge theory correlators from non-critical string theory,''
  Phys.\ Lett.\ B {\bf 428}, 105 (1998)
  [arXiv:hep-th/9802109].
  
  \bibitem{Witten:1998qj}
  E.~Witten,
  ``Anti-de Sitter space and holography,''
  Adv.\ Theor.\ Math.\ Phys.\  {\bf 2}, 253 (1998)
  [arXiv:hep-th/9802150].
    
   \bibitem{Brodsky:2008pg}
  S.~J.~Brodsky and G.~F.~de Teramond,
  ``AdS/CFT and Light-Front QCD,''
  arXiv:0802.0514 [hep-ph],
  and references therein.
  
  \bibitem{Deur:2008rf}
  A.~Deur, V.~Burkert, J.~P.~Chen and W.~Korsch,
  ``Determination of the effective strong coupling constant $\alpha_{s, g_1}(Q^2)$
  from CLAS spin structure function data,''
  arXiv:0803.4119 [hep-ph].
    
  \bibitem{Cornwall:1981zr}
  J.~M.~Cornwall,
  ``Dynamical Mass Generation In Continuum QCD,''
  Phys.\ Rev.\  D {\bf 26}, 1453 (1982).

  \bibitem{Parisi:1972zy}
  G.~Parisi,
  ``Conformal invariance in perturbation theory,''
  Phys.\ Lett.\  B {\bf 39}, 643 (1972).
  
   \bibitem{Brodsky:1985ve}
  S.~J.~Brodsky, Y.~Frishman and G.~P.~Lepage,
  ``On The Application Of Conformal Symmetry To Quantum Field Theory,''
  Phys.\ Lett.\  B {\bf 167}, 347 (1986).
  
   \bibitem{Polchinski:2001tt}
  J.~Polchinski and M.~J.~Strassler,
  ``Hard scattering and gauge/string duality,''
  Phys.\ Rev.\ Lett.\  {\bf 88}, 031601 (2002)
  [arXiv:hep-th/0109174].
  
 \bibitem{Karch:2006pv}
  A.~Karch, E.~Katz, D.~T.~Son and M.~A.~Stephanov,
  ``Linear confinement and AdS/QCD,''
  Phys.\ Rev.\  D {\bf 74}, 015005 (2006)
  [arXiv:hep-ph/0602229].
  
   \bibitem{Brodsky:2003px}
  S.~J.~Brodsky and G.~F.~de Teramond,
  ``Light-front hadron dynamics and AdS/CFT correspondence,''
  Phys.\ Lett.\  B {\bf 582}, 211 (2004)
  [arXiv:hep-th/0310227].
  
  \bibitem{Boschi-Filho:2002vd}
  H.~Boschi-Filho and N.~R.~F.~Braga,
  ``Gauge / string duality and scalar glueball mass ratios,''
  JHEP {\bf 0305}, 009 (2003)
  [arXiv:hep-th/0212207].
  
  \bibitem{deTeramond:2005su}
  G.~F.~de Teramond and S.~J.~Brodsky,
  ``The hadronic spectrum of a holographic dual of QCD,''
  Phys.\ Rev.\ Lett.\  {\bf 94}, 201601 (2005)
  [arXiv:hep-th/0501022].
  
   \bibitem{Erlich:2005qh}
  J.~Erlich, E.~Katz, D.~T.~Son and M.~A.~Stephanov,
  ``QCD and a holographic model of hadrons,''
  Phys.\ Rev.\ Lett.\  {\bf 95}, 261602 (2005)
  [arXiv:hep-ph/0501128].

  \bibitem{DaRold:2005zs}
  L.~Da Rold and A.~Pomarol,
  ``Chiral symmetry breaking from five dimensional spaces,''
  Nucl.\ Phys.\  B {\bf 721}, 79 (2005)
  [arXiv:hep-ph/0501218].
  
  
  \bibitem{Brodsky:2008pf}
  S.~J.~Brodsky and G.~F.~de Teramond,
  ``Light-Front Dynamics and AdS/QCD Correspondence: Gravitational Form Factors
  of Composite Hadrons,''
  arXiv:0804.0452 [hep-ph].

   \bibitem{Brodsky:1997de}
  S.~J.~Brodsky, H.~C.~Pauli and S.~S.~Pinsky,
  ``Quantum chromodynamics and other field theories on the light cone,''
  Phys.\ Rept.\  {\bf 301}, 299 (1998)
  [arXiv:hep-ph/9705477].
  
   \bibitem{Brodsky:2000ii}
  S.~J.~Brodsky, D.~S.~Hwang, B.~Q.~Ma and I.~Schmidt,
  ``Light-cone representation of the spin and orbital angular momentum of
  relativistic composite systems,''
  Nucl.\ Phys.\  B {\bf 593}, 311 (2001)
  [arXiv:hep-th/0003082].
  
   \bibitem{Soper:1976jc}
  D.~E.~Soper,
  ``The Parton Model And The Bethe-Salpeter Wave Function,''
  Phys.\ Rev.\ D {\bf 15}, 1141 (1977).
  
 \bibitem{Ji:1996ek}
  X.~D.~Ji,
  ``Gauge invariant decomposition of nucleon spin,''
  Phys.\ Rev.\ Lett.\  {\bf 78}, 610 (1997)
  [arXiv:hep-ph/9603249].

  
 \bibitem{Teryaev:1999su}
  O.~V.~Teryaev,
  ``Spin structure of nucleon and equivalence principle,''
  arXiv:hep-ph/9904376.

   \bibitem{Abidin:2008ku}
  Z.~Abidin and C.~E.~Carlson,
  ``Gravitational Form Factors of Vector Mesons in an AdS/QCD Model,''
  arXiv:0801.3839 [hep-ph];
  ``Gravitational Form Factors in the Axial Sector from an AdS/QCD Model,''
  arXiv:0804.0214 [hep-ph].
  
  \bibitem{Breitenlohner:1982jf}
  P.~Breitenlohner and D.~Z.~Freedman,
  ``Stability In Gauged Extended Supergravity,''
  Annals Phys.\  {\bf 144}, 249 (1982).
  
  \bibitem{l'Yi:1998eu}
  W.~S.~l'Yi,
  ``Correlators of currents corresponding to the massive p-form fields in
  AdS/CFT correspondence,''
  Phys.\ Lett.\ B {\bf 448}, 218 (1999)
  [arXiv:hep-th/9811097].

  
    \bibitem{Lepage:1979zb}
  G.~P.~Lepage and S.~J.~Brodsky,
  ``Exclusive Processes In Quantum Chromodynamics: Evolution Equations For
  Hadronic Wave Functions And The Form-Factors Of Mesons,''
  Phys.\ Lett.\  B {\bf 87}, 359 (1979).
  
  \bibitem{Aitala:2000hb}
  E.~M.~Aitala {\it et al.}  [E791 Collaboration],
  ``Direct measurement of the pion valence quark momentum distribution, the
  pion light-cone wave function squared,''
  Phys.\ Rev.\ Lett.\  {\bf 86}, 4768 (2001)
  [arXiv:hep-ex/0010043].
  
  \bibitem{Choi:2006ha}
  H.~M.~Choi and C.~R.~Ji,
  ``Conformal Symmetry and Pion Form Factor: Soft and Hard Contributions,''
  Phys.\ Rev.\  D {\bf 74}, 093010 (2006)
  [arXiv:hep-ph/0608148].
    
\bibitem{Brodsky:1973kb}
  S.~J.~Brodsky, R.~Roskies and R.~Suaya,
  ``Quantum Electrodynamics And Renormalization Theory In The Infinite Momentum
  Frame,''
  Phys.\ Rev.\  D {\bf 8}, 4574 (1973).

\bibitem{Munger:1993kq}
  C.~T.~Munger, S.~J.~Brodsky and I.~Schmidt,
  ``Production of relativistic anti-hydrogen atoms by pair production with
  positron capture,''
  Phys.\ Rev.\  D {\bf 49}, 3228 (1994).

\bibitem{White:1994tj}
  C.~G.~White {\it et al.},
  ``Comparison of 20 exclusive reactions at large t,''
  Phys.\ Rev.\  D {\bf 49}, 58 (1994).
  
    
\bibitem{Brodsky:2008xm}
  S.~J.~Brodsky and R.~Shrock,
  ``On Condensates in Strongly Coupled Gauge Theories,''
  arXiv:0803.2541 [hep-th];
  ``Standard-Model Condensates and the Cosmological Constant,''
  arXiv:0803.2554 [hep-th].





\end{thebibliography}
\end{document}